\documentclass{PoS}%
\usepackage{amsmath}
\usepackage{amsfonts}
\usepackage{amssymb}
\usepackage{graphicx}%
\providecommand{\U}[1]{\protect\rule{.1in}{.1in}}
\input{frontmatter.text}
\begin{document}
\section{Introduction}

Quark confinement follows from the area law of the Wilson loop average. The
dual Meissner effect is the promising mechanism for quark confinement
\cite{dualSC}. Based on the Abelian projection there are many numerical
analyses which support the dual superconductivity picture, such as Abelian
dominance \cite{Suzuki90}, magnetic monopole dominance \cite{stack94}
\cite{shiba}, and center vortex dominance \cite{greensite} in the string
tension. However, these results are obtained only in special gauges such as
the maximal Abelian (MA) gauge and the Laplacian Abelian gauge. Such gauge
fixings also cause the problem of the color symmetry (global symmetry) breaking.

\begin{figure}[ptb]
\begin{center}
\includegraphics[
height=6cm,
angle=0
]
{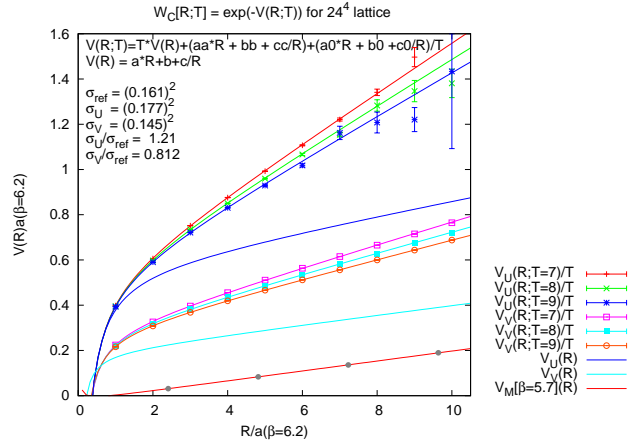}
\end{center}
\caption{{}The quark-antiquark potential calculated using three types of
sources shown from top to bottom: (a) the original YM field ($U$), (b) the
restricted $U(2)$ field ($V$-field) and (c) the non-abelian magnetic
monopole.}%
\label{fig:potential}%
\end{figure}

To overcome these problems, we have presented a new lattice formulation of
$SU(N)$ Yang-Mills (YM) theory\cite{KSM05}\cite{SCGTKKS08}, that gives the
decomposition of the gauge link variable suited for extracting the dominant
mode for quark confinement. In the case of the $SU(2)$ YM theory, the
decomposition of the gauge link variable is given by a compact representation
of Cho-Duan-Ge-Faddeev-Niemi (CDGFN) decomposition\cite{CFNS-C} on a
lattice.\cite{ref:NLCVsu2}\cite{ref:NLCVsu2-2} For the $SU(N)$ YM theory, the
new formula for the decomposition of the gauge link variable is constructed as
an extension of the $SU(2)$ case. There are several possibilities of
decomposition an extension of the $SU(2)$ case. There are several
possibilities of decomposition corresponding to the stability subgroup
$\tilde{H}$ of gauge symmetry group $G,$ while there is the unique option of
$\tilde{H}=U(1)$ in the $SU(2).$ For the case of $G=SU(3)$, there are two
possibility which we call the maximal option and the minimal option. The
maximal option is obtained for the stability group $\tilde{H}=U(1)\times
U(1)$, which is the gauge invariant version of the Abelian projection in the
maximal Abelian (MA) gauge \cite{lattce2007}\cite{lattice2008}. The minimal
one is obtained for $\tilde{H}=U(2)\cong SU(2)\times U(1)$, suitable for the
Wilson loop in the fundamental representation \cite{KondoNAST}. We have
demonstrated the gauge independent (invariant) restricted $U(2)$-dominance,
(or conventionally called "Abelian\textquotedblright\ dominance), $(\sigma
_{V}/\sigma_{full}=93\pm16\%)$ where the decomposed $V$-field (restricted U(2)
field) reproduced the string tension of original YM field, and \ the gauge
independent non-Abelian magnetic monopole dominance ($\sigma_{mon}/\sigma
_{V}=94\pm9\%$), where the string tension was reproduced by only the
(non-Abelian) magnetic monopole part extracted from the restricted U(2) field
(see Figure \ref{fig:potential})\cite{lattice2009}\cite{kato:lattice2009}%
\cite{lattice2010}\cite{abeliandomSU(3)}.

To establish the dual superconductivity picture, the dual Meissner effect in
Yang-Mills theory must be examined by measuring the distribution of
chromo-electric field strength or color flux created by a static
quark-antiquark pair. There are many works on color flux for the Yang-Mills
field by using Wilson line/loop operator\cite{Cardaci2011} \cite{Bakry}
\cite{Cardso} \cite{Bicudo} \cite{Cea}. However, there is no direct
measurement of the dual Meissner effect for the $SU(3)$ YM theory in the gauge
independent (invariant) way. In this talk, we focus on the dual Meissner
effect. Applying our new formulation to the $SU(3)$ YM theory, we measure the
distribution of the extracted chromo-electric field strength which is created
by a static quark-antiquark pair. Then, we investigate whether or not the
non-Abelian dual superconductivity claimed by us is indeed a mechanism of
quark confinement in $SU(3)$ YM theory.

\section{Extracting the dominant mode for dual super conductivity}

We briefly review a new formulation of the lattice YM theory, which extracts
non-Abelian dual superconductivity mode for $SU(3)$ YM
theory.\cite{abeliandomSU(3),lattice2010} In this talk, we use the minimal
option, since we consider the quark confinement in the fundamental
representation. Let us consider the decomposition of link variables,
$U_{x,\mu}=X_{x,\mu}V_{x,\mu}$, where $V_{x.\mu}$ \ could be the dominant mode
for quark confinement, and $X_{x,\mu}$ the remainder part. The YM field and
the decomposed new-variables are transformed by full $SU(3)$ gauge
transformation $\Omega_{x}$ such that $V_{x,\mu}$ is transformed as the gauge
link variable and $X_{x,\mu}$ as the site available:
\begin{subequations}
\label{eq:gaugeTransf}%
\begin{align}
U_{x,\mu}  &  \longrightarrow U_{x,\nu}^{\prime}=\Omega_{x}U_{x,\mu}%
\Omega_{x+\mu}^{\dag},\\
V_{x,\mu}  &  \longrightarrow V_{x,\nu}^{\prime}=\Omega_{x}V_{x,\mu}%
\Omega_{x+\mu}^{\dag},\text{ \ }X_{x,\mu}\longrightarrow X_{x,\nu}^{\prime
}=\Omega_{x}X_{x,\mu}\Omega_{x}^{\dag}.
\end{align}
The decomposition is determined by the defining equation
\end{subequations}
\begin{subequations}
\begin{align}
&  D_{\mu}^{\epsilon}[V]\mathbf{h}_{x}:=\frac{1}{\epsilon}\left[  V_{x,\mu
}\mathbf{h}_{x+\mu}-\mathbf{h}_{x}V_{x,\mu}\right]  =0,\label{eq:def1}\\
&  g_{x}:=e^{i2\pi q/N}\exp(-ia_{x}^{0}\mathbf{h}_{x}-i\sum\nolimits_{j=1}%
^{3}a_{x}^{(j)}\mathbf{u}_{x}^{(i)})=1, \label{eq:def2}%
\end{align}
where $\mathbf{h}_{x}$ is a introduced color field $\mathbf{h}_{x}=\xi
(\lambda^{8}/2)\xi^{\dag}$ $\in\lbrack SU(3)/U(2)]$ with $\lambda^{8}$ being
the Gell-Mann matrix and $\xi$ the $SU(3)$ gauge element. The variable $g_{x}$
is undetermined parameter from Eq.(\ref{eq:def1}), $\mathbf{u}_{x}^{(j)}$ 's
are $su(2)$-Lie algebra values\thinspace, and $q_{x}$ an integer value
$\ 0,1,2,...,N-1$, respectively. These defining equations can be solved
exactly \cite{exactdecomp}, and the solution is given by
\end{subequations}
\begin{subequations}
\label{eq:decomp}%
\begin{align}
L_{x,\mu}  &  =\frac{N^{2}-2N+2}{N}\mathbf{1}+(N-2)\sqrt{\frac{2(N-1)}{N}%
}(\mathbf{h}_{x}+U_{x,\mu}\mathbf{h}_{x+\mu}U_{x,\mu}^{\dag}),\nonumber\\
&  +4(N-1)\mathbf{h}_{x}U_{x,\mu}\mathbf{h}_{x+\mu}U_{x,\mu}^{\dag},\\
\widehat{L}_{x,\mu}  &  =\left(  L_{x,\mu}L_{x,\mu}^{\dag}\right)
^{-1/2}L_{x,\mu},\\
X_{x,\mu}  &  =\widehat{L}_{x,\mu}^{\dag}\det(\widehat{L}_{x,\mu})^{1/N}%
g_{x}^{-1},\text{ \ \ \ }V_{x,\mu}=X_{x,\mu}^{\dag}U_{x,\mu}=g_{x}\widehat
{L}_{x,\mu}U_{x,\mu}.
\end{align}
Note that the above defining equations correspond to the continuum version:
$D_{\mu}[\mathcal{V}]\mathbf{h}(x)=0$ and $\mathrm{tr}(\mathbf{h}%
(x)\mathcal{X}_{\mu}(x))=0,$ respectively. In the naive continuum limit, we
have the corresponding decomposition, $\mathbf{A}_{\mathbf{\mu}}%
(x)=\mathbf{V}_{\mu}(x)+\mathbf{X}_{\mu}(x)$ as
\end{subequations}
\begin{subequations}
\begin{align}
\mathbf{V}_{\mu}(x)  &  =\mathbf{A}_{\mathbf{\mu}}(x)-\frac{2(N-1)}{N}\left[
\mathbf{h}(x),\left[  \mathbf{h}(x),\mathbf{A}_{\mathbf{\mu}}(x)\right]
\right]  -ig^{-1}\frac{2(N-1)}{N}\left[  \partial_{\mu}\mathbf{h}%
(x),\mathbf{h}(x)\right]  ,\\
\mathbf{X}_{\mu}(x)  &  =\frac{2(N-1)}{N}\left[  \mathbf{h}(x),\left[
\mathbf{h}(x),\mathbf{A}_{\mathbf{\mu}}(x)\right]  \right]  +ig^{-1}%
\frac{2(N-1)}{N}\left[  \partial_{\mu}\mathbf{h}(x),\mathbf{h}(x)\right]  .
\end{align}

To determine the decomposition, we need to determine the color field. Here, we
use the reduction condition to minimize the functional
\end{subequations}
\begin{equation}
F_{\text{red}}[\mathbf{h}_{x}]=\sum_{x,\mu}\mathrm{tr}\left\{  (D_{\mu
}^{\epsilon}[U_{x,\mu}]\mathbf{h}_{x})^{\dag}(D_{\mu}^{\epsilon}[U_{x,\mu
}]\mathbf{h}_{x})\right\}  , \label{eq:reduction}%
\end{equation}
which is equivalent to solving the ground state of a spin glass.

\begin{figure}[ptb]
\begin{center}
\includegraphics[
height=5cm,
]
{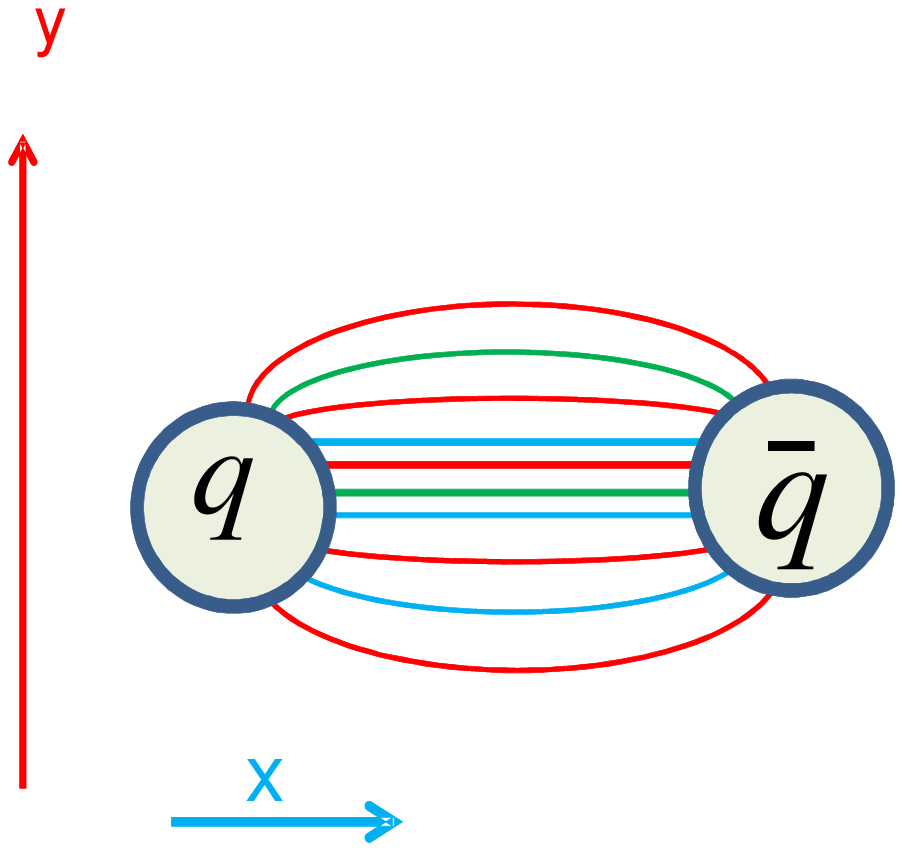} \ \ \ \ \ \ \ \ \ \ \includegraphics[
origin=c,
height=5.5cm,
angle=270
]
{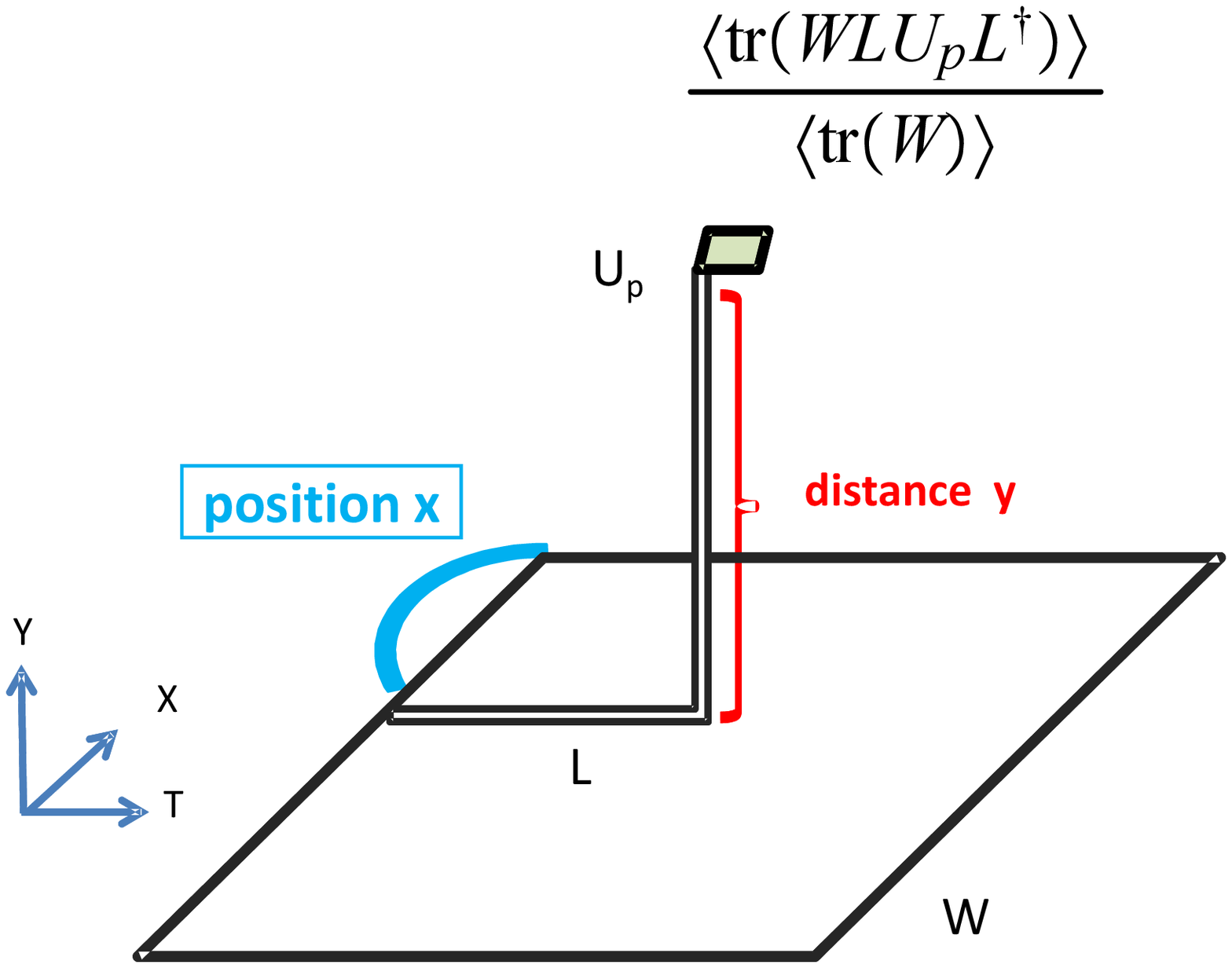}
\end{center}
\caption{(Left) The measurement setup of the color flux produced by a
quark--antiquark pair. (Right) The connected correlator between a plaquette
and the Wilson loop.}%
\label{Fig:Operator}%
\end{figure}

From the non-Abelian Stokes' theorem\cite{KondoNAST}\cite{KondoShibata} and
the Hodge decomposition of the field strength $\mathcal{F}_{\mu\nu}%
[\mathbf{V}],$ we obtain the magnetic part (magnetic monopole current) in the
gauge independent way. Therefore, the gauge invariant magnetic monopole is
defined by using the $V$-field as
\begin{subequations}
\begin{align}
&  V_{x,\mu}V_{x+\mu,\mu}V_{x+\nu,\mu}^{\dag}V_{x,\nu}^{\dag}=\exp\left(
-ig\mathcal{F}_{\mu\nu}[\mathbf{V}]\right)  =\exp(-ig\Theta_{\mu\nu}%
^{8}\mathbf{h}(x))\\
&  \Theta_{\mu\nu}^{8}:=-\arg\text{ \textrm{Tr}}\left[  \left(  \frac{1}%
{3}\mathbf{1}-\frac{2}{\sqrt{3}}\mathbf{h}_{x}\right)  V_{x,\mu}V_{x+\mu,\mu
}V_{x+\nu,\mu}^{\dag}V_{x,\nu}^{\dag}\right]  ,\\
&  k_{\mu}=2\pi n_{\mu}:=\frac{1}{2}\epsilon_{\mu\nu\alpha\beta}\partial_{\nu
}\Theta_{\alpha\beta}^{8}.
\end{align}
Note that the the current $k_{\mu}$ is the non-Abelian magnetic monopole
current, since $V_{x,\mu}$ is the decomposed restricted $U(2)$ filed.

\section{Measurements of color flux}

We investigate the color flux produced by a quark-antiquark pair, see the left
panel of Fig. \ref{Fig:Operator}. In order to explore the color flux in the
gauge invariant way, we use the connected correlator of the Wilson line
\cite{DiGiacomo} (see the right panel\ of Fig.\ref{Fig:Operator}),%
\end{subequations}
\begin{equation}
\rho_{W}:=\frac{\left\langle \mathrm{tr}\left(  U_{p}L^{\dag}WL\right)
\right\rangle }{\left\langle \mathrm{tr}\left(  W\right)  \right\rangle
}-\frac{1}{N}\frac{\left\langle \mathrm{tr}\left(  U_{p}\right)
\mathrm{tr}\left(  W\right)  \right\rangle }{\left\langle \mathrm{tr}\left(
W\right)  \right\rangle }, \label{eq:Op}%
\end{equation}
where $W\,\ $represents the Wilson loop as the quark and anti-quark source
term, $U_{p}$ a plaquette variable as the probe operator to measure field
strength, $L$ the Wilson line connecting the source $W$ and probe $U_{p}$, and
$N$ the number of color ($N=3$). The symbol $\left\langle \mathcal{O}%
\right\rangle $ denotes the average of the operator $\mathcal{O}$ in the space
and the ensemble of the configurations. Note that this is sensitive to the
field strength rather than the disconnected one. Indeed, in the naive
continuum limit, the connected correlator $\rho_{W}$ is given by
\begin{equation}
\rho_{W}\overset{\varepsilon\rightarrow0}{\simeq}g\epsilon^{2}\left\langle
\mathcal{F}_{\mu\nu}\right\rangle _{q\bar{q}}:=\frac{\left\langle
\mathrm{tr}\left(  g\epsilon^{2}\mathcal{F}_{\mu\nu}L^{\dag}WL\right)
\right\rangle }{\left\langle \mathrm{tr}\left(  W\right)  \right\rangle
}+O(\epsilon^{4}),
\end{equation}
while the disconnected one is given by
\begin{equation}
\rho_{W}^{\prime}:=\frac{\left\langle \mathrm{tr}\left(  W\right)
\mathrm{tr}\left(  U_{p}\right)  \right\rangle }{\left\langle \mathrm{tr}%
\left(  W\right)  \right\rangle }-\left\langle \mathrm{tr}\left(
U_{p}\right)  \right\rangle \overset{\varepsilon\rightarrow0}{\simeq}%
g\epsilon^{4}\left[  \left\langle \mathcal{F}_{\mu\nu}^{2}\right\rangle
_{q\bar{q}}-\left\langle \mathcal{F}_{\mu\nu}^{2}\right\rangle _{0}\right]  .
\end{equation}
Thus, the color filed strength is given by $\ F_{\mu\nu}=\sqrt{\frac{\beta
}{2N}}\rho_{W}$.

To investigate the non-Abelian dual Meissner effect as the mechanism of quark
confinement, we measure correlators of the restricted $U(2)$ field, $V_{x,\mu
}$, in place of the original YM filed. Note again that this restricted
$U(2)$-field and the non-Abelian magnetic monopole extracted from it reproduce
the string tension in the quark--antiquark potential. (see Fig.
\ref{fig:potential})

We generate gauge configurations using the standard Wilson action on a
$24^{4}$ lattice with $\beta=6$. The gauge link decomposition is obtained by
the formula given in the previous section, i.e., the color field configuration
is obtained by solving the reduction condition of minimizing the functional
eq(\ref{eq:reduction}), and the decomposition is obtained by using the formula
eq(\ref{eq:decomp}). In the measurement, we apply the APE smearing method to
reduce the noise.

\begin{figure}[ptb]
\begin{center}
\includegraphics[
height=7cm,
angle=270
]
{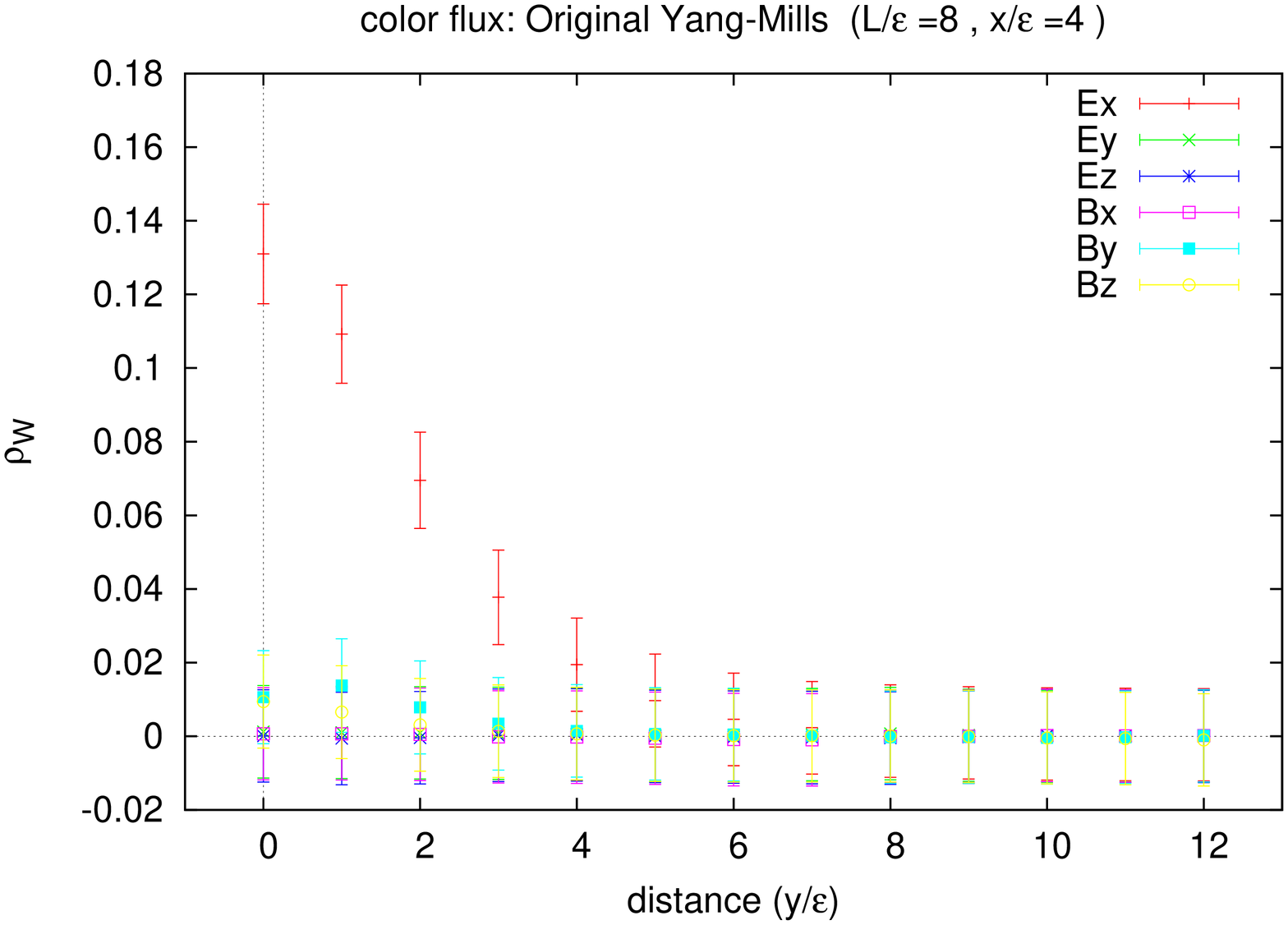} \includegraphics[
height=7cm,
angle=270
]
{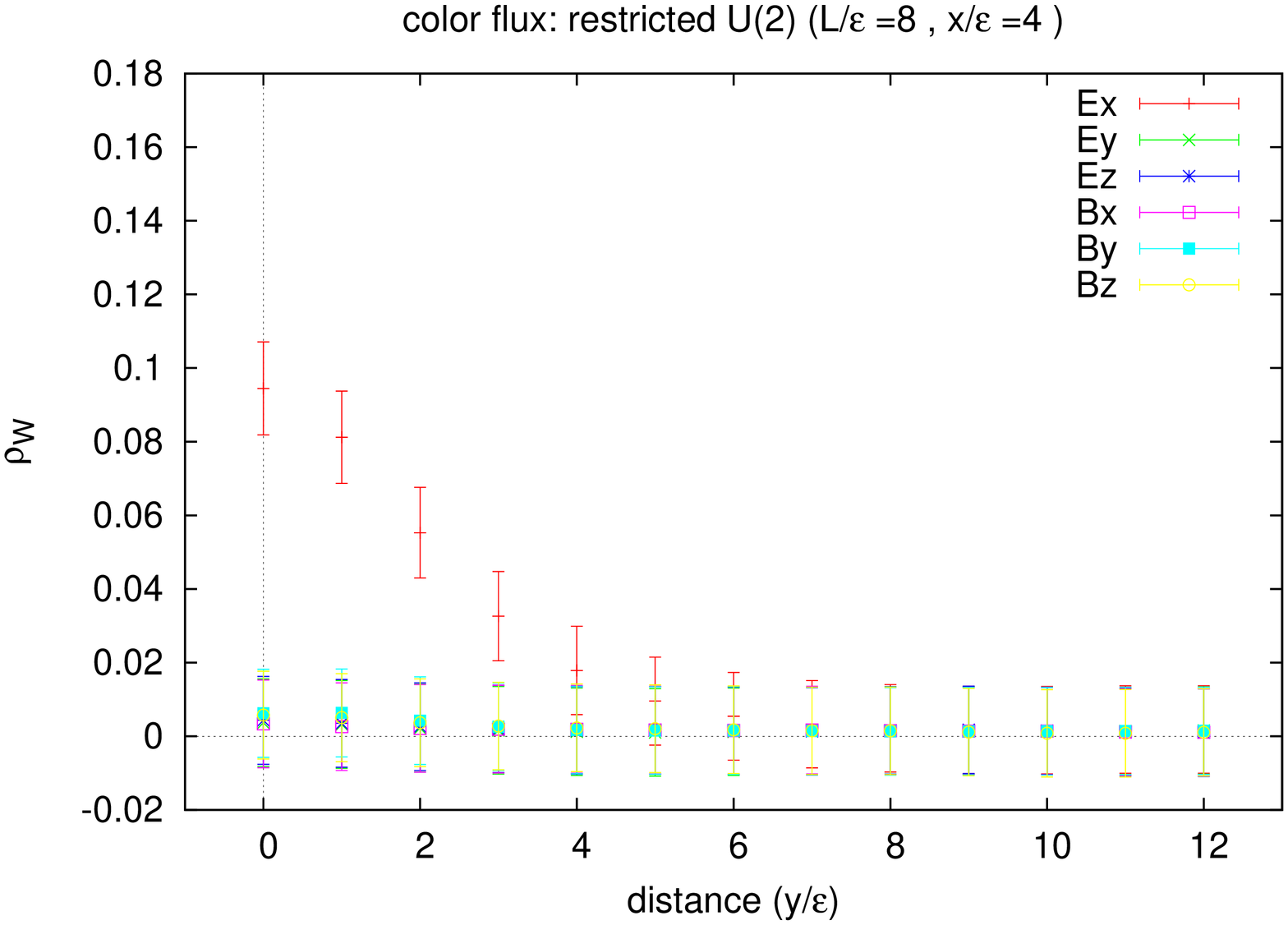}
\end{center}
\caption{Measurement of $E_{x}$ component of the color flux in the x-y plane.
(Left panel) \ The $E_{x}$ for the $SU(3)$ YM field (Right panel) \ $E_{x}$
component for the restricted $U(2)$ field.}%
\label{fig:measure1}%
\end{figure}

\begin{figure}[ptb]
\begin{center}
\includegraphics[
height=7cm,
angle=270
]
{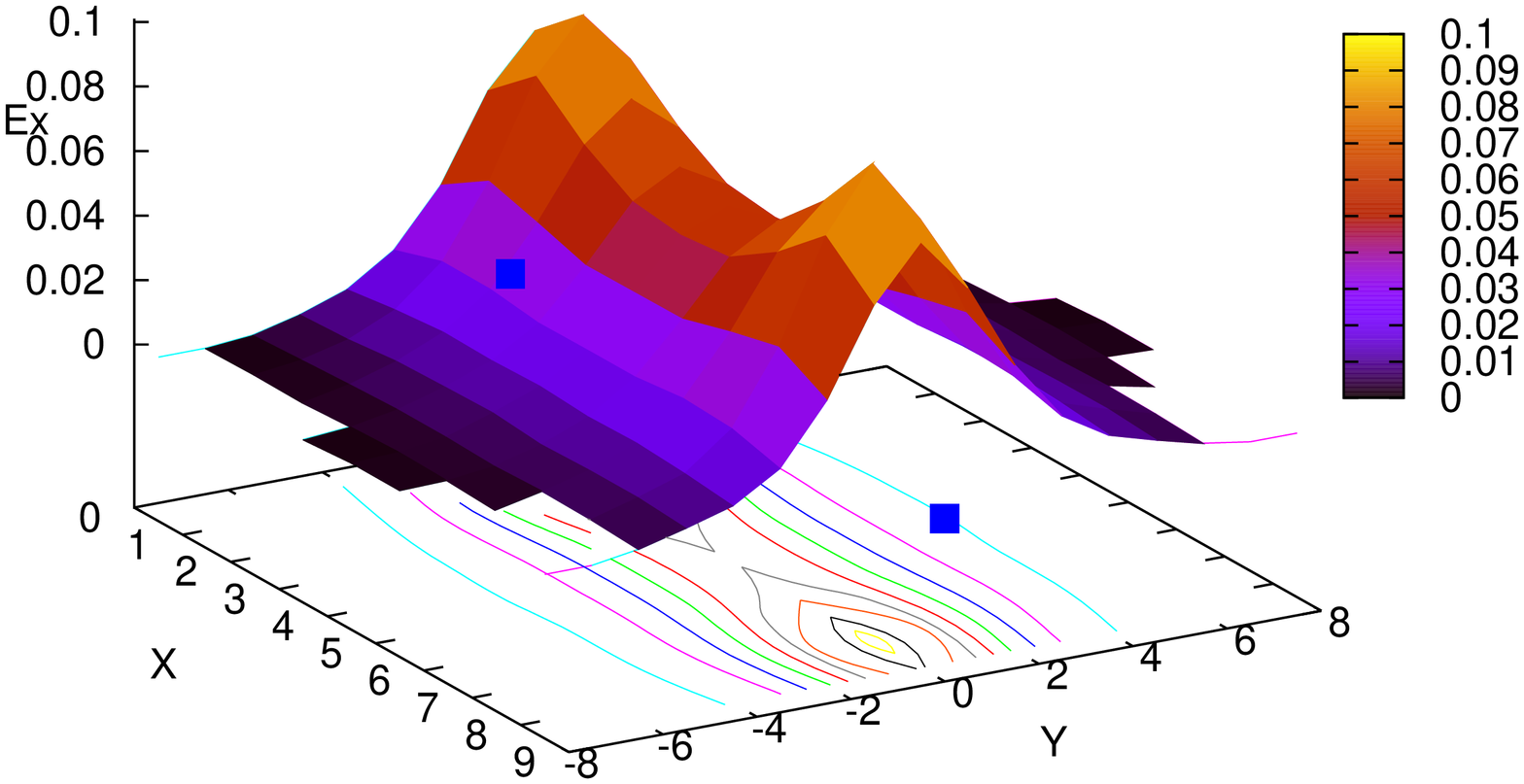} \includegraphics[
height=7cm,
angle=270
]
{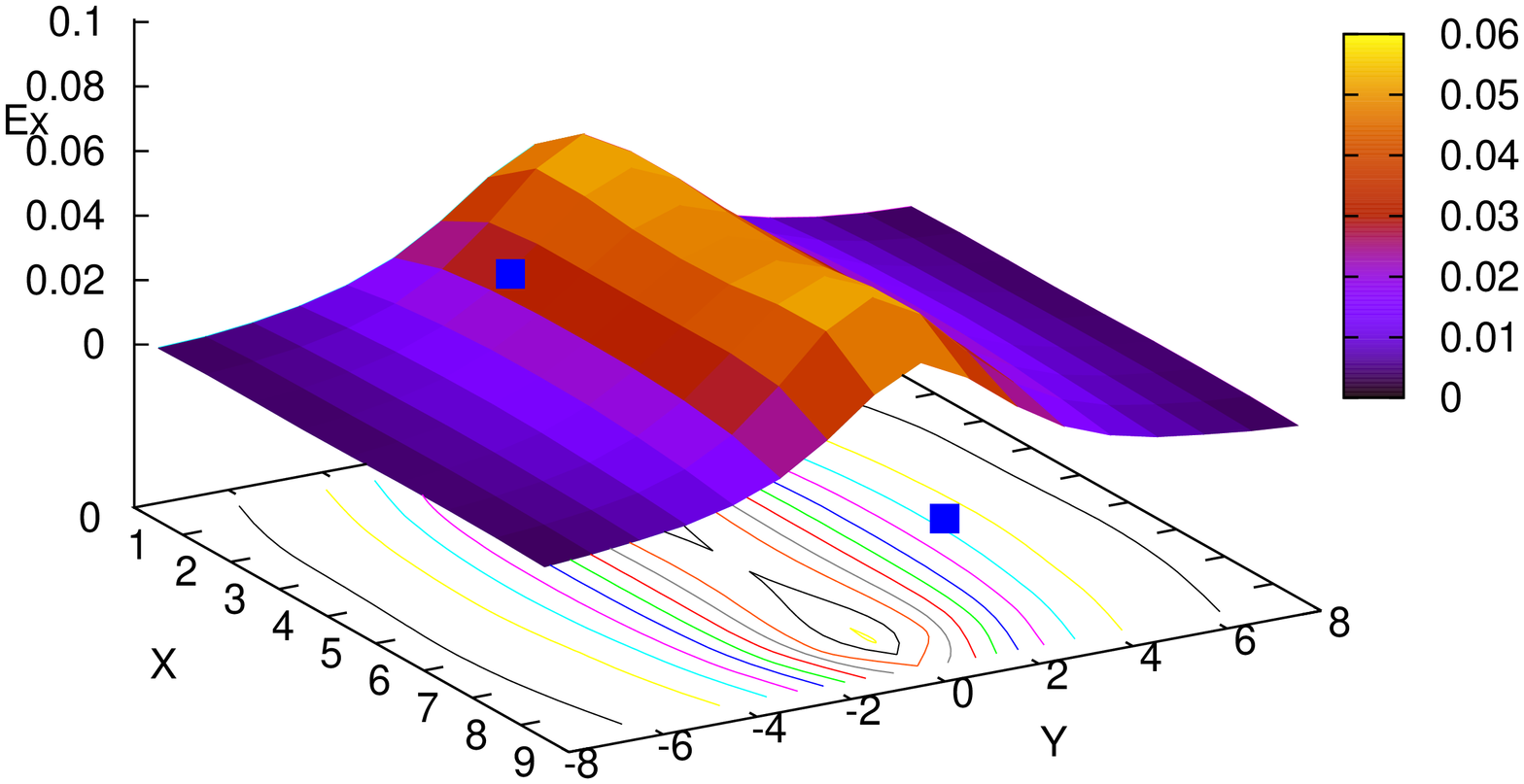}
\end{center}
\caption{Measurement \ of color flux. (Left panel) \ $E_{x}$ of the $SU(3)$ YM
field. (Right panel) \ $E_{x}$ component of the restricted $U(2)$ field.}%
\label{fig:fluxtube}%
\end{figure}

The left panel of Fig. \ref{fig:measure1} shows the result of measurements of
the color flux obtained from the original YM field. Here the quark and
antiquark source is introduce as $8\times8$ Wilson loop ($W$) in the X-T
plane, and the probe $(U_{p})$ is set at the center of Wilson loop and moved
along the Y-direction. (see Fig. \ref{Fig:Operator}). The $E_{x}$ component
only has non-zero value, and it decreases quickly as away from the Wilson
loop. The right panel of Fig. \ref{fig:measure1} shows the color flux obtained
from the restricted $U(2)$-field, where the gauge link variable $V_{\,x,\mu}%
$\ is used in place of the YM field $U_{x,\mu}$ in eq(\ref{eq:Op}). The
$E_{x}$ component only has non-zero value as well as the original YM filed
case, and color flux is detected only along the Wilson loop. Then, to
investigate the profile of the color flux tube we explore the distribution of
color flux in the dimensional plane. Fig. \ref{fig:fluxtube} shows the
magnitude of $E_{x}$ of the color flux in the X-Y plane. The quark and
antiquark source is introduced as $9\times11$ Wilson loop in the X-T plane.
The probe is displaced on the X-Y plane at the midpoint of the Wilson loop in
the T-direction. The position of the quark and antiquark is marked by the
solid blue box. The magnitude is shown by the height of the 3D plot and also
the contour in the bottom plane. The left panel shows the color flux of the YM
filed, and the right panel of the restricted $U(2)$-field. The figures show
that the color flux of the restricted $U(2)$-field or $\mathbf{V}_{\mu}(x)$
component, is narrowed down to the tube. While, as for the original YM field
the color flux of the $\mathbf{X}_{\mu}(x)$ component could be detected near
the quark and antiquark sources.

\section{Summary and outlook}

We have studied the dual Meissner effect for $SU(3)$ YM theory by using our
new formulation of YM theory on a lattice. We have extracted the restricted
$U(2)$ field from the YM field which play a dominant role in confinement of
quark (fermion in the fundamental representation), i.e., the restricted $U(2)$
dominance and the non-Abelian magnetic monopole dominance in the string
tension. By measuring the color flux for both the original YM field and the
restricted $U(2)$ field, we have investigated the dual Messier effect of
$SU(3)$ YM theory. We have found that the electric part of the color flux is
narrowed down in both cases, and the restricted $U(2)$ field part is dominant
in the long distance region. We have found the tube-shaped flux of the
electric part of the restricted $U(2)$ color field, which is a direct evidence
of the non-Abelian dual superconductivity.

To confirm the non-Abelian dual superconductivity picture, we need further
study on the dual Meissner effect of $SU(3)$ YM theory, e.g., determination of
the type of dual super conductor, measurement of the penetrating depth,
induced magnetic current around color flux due to the magnetic monopole
condensation, and so on.

\subsection*{Acknowledgement}

This work is supported by Grant-in-Aid for Scientific Research (C) 21540256
from Japan Society for the Promotion Science (JSPS), and also in part by JSPS
Grant-in-Aid for Scientific Research (S) 22224003. The numerical calculations
are supported by the Large Scale Simulation Program No.09/10-19 (FY2009-2010)
and No.10-13 (FY2010) of High Energy Accelerator Research Organization (KEK).

\end{document}